\documentclass[aps,prl,twocolumn,showpacs,preprintnumbers,amsmath,amssymb]{revtex4}
\usepackage{graphicx}
\usepackage{times}

\newcommand{\beq}{\begin{eqnarray}}   \newcommand{\eeq}{\end{eqnarray}}
\newcommand{\nn}{\nonumber}
\newcommand{\bra}{\langle}  \newcommand{\ket}{\rangle}
\newcommand{\del}{\partial} 

  \newcommand{\tr}{\mbox{tr}}
  
\newcommand{\Nred}{N_{\rm red}}

\newcommand{\Tpc}{T_{\rm pc}}

\begin{document}

\title{On early onset of quark number density at zero temperature}


\author{Keitaro Nagata}
\affiliation{
{\textit RIISE, Hiroshima University, Higashi-Hiroshima 739-8527 JAPAN}
}
\email[]{kngt@hiroshima-u.ac.jp}

\date{\today}

\begin{abstract}
We study a longstanding problem in lattice QCD at low temperature and nonzero 
quark chemical potential on an onset of the quark number density 
at $\mu=m_\pi/2$.
We introduce a physical parametrization of the eigenvalues in 
the reduction formula of the fermion determinant.
It is shown that the parametrization reduces the quark number density 
operator to an expression with the Fermi distribution of the quark. 
For each configuration, the eigenvalues of the reduced matrix 
correspond to one-particle energy states of a quark. 
The gap of the eigenspectrum of the reduced matrix corresponds to the 
gap of the energy states, which causes the $\mu$-independence 
of the fermion determinant for small $\mu$ at $T=0$. 
Once $\mu$ exceeds the gap, the quark number density 
becomes nonzero for each configuration, 
which causes the early onset of the quark number density. 
\end{abstract}


\maketitle

Properties of QCD at finite temperature ($T$) and quark chemical potential 
($\mu$) are central issues in particle, nuclear and astrophysics in order 
to understand various states of matters in our universe. 
One important nature of QCD is the deconfinement transition. 
Recent simulations on finer lattices show that the deconfinement transition 
is crossover~\cite{Aoki:2006we}, and occurs at $\Tpc=150-170$ MeV depending 
on observables~\cite{Borsanyi:2010bp,Borsanyi:2010cj,Bazavov:2012ty}.
It is expected that the crossover extends to nonzero $\mu$ and turns 
into a first order phase transition at a critical endpoint(CEP), which
has been studied in many references, e.g.~\cite{Stephanov:2007fk,Philipsen:2008gf}. 

The QCD phase diagram is expected to contain rich structures also 
at low temperatures~\cite{Fukushima:2010bq,Stephanov:2007fk,Ohnishi:2011aa}.
In addition, the discovery of a pulser with twice solar 
mass~\cite{Demorest:2010bx} calls the reliable equation of state of QCD. 
Properties of QCD at low $T$ and nonzero $\mu$ was investigated in 
lattice simulations with a Glasgow method~\cite{Barbour:1999mc,Barbour:1997ej,Barbour:1997bh,Barbour:1991vs}, 
where they obtained a signal of a finite density phase transition. 
They also obtained an onset of the quark number density at $\mu=m_\pi/2$, 
which disagrees with a phenomenological expectation that the baryon number 
density remains zero for $\mu_B = 3\mu < M_N$. 
Although several attempts were made, e.g. ~\cite{Aloisio:1998cn,Crompton:2001ws}, the solution is not yet obtained. 
In order to reveal the QCD phase diagram based on first principle 
lattice simulations, one must confront this problem. 

It was shown theoretically~\cite{Cohen:2003kd,Adams:2004yy}
that the fermion determinant $\det \Delta(\mu)$ is independent 
of $\mu$ at $T=0$. 
Cohen showed the $\mu$-independence for isospin chemical potential case 
by considering the $\gamma_0$ times the Dirac operator. 
Adams showed it for quark chemical potential case 
by using a reduction formula of the fermion determinant. 
For lattice QCD, we showed the $\mu$-independence of $\det \Delta(\mu)$ 
by using the reduction formula~\cite{Nagata:2012tc} in two approaches. 
One approach was to evaluate the $\mu$- and $T$-dependence of 
a factor $\det \Delta(\mu)/\det \Delta(0)$. 
We found that the factor becomes insensitive to $\mu$ as $T$ decreases. 
The other one was to take the low temperature limit 
of $\det \Delta(\mu)$ with the aid of properties of the 
reduced matrix, a $N_t$ scaling law, the existence of a gap, pair nature and 
a relation between the gap and pion mass. 
We found that those properties makes $\det \Delta(\mu)$ $\mu$-independent for $\mu<m_\pi/2$. 

Thus, the $\mu$-independence of the fermion determinant at low $T$ 
and small $\mu$ was explained both in theoretical studies and lattice 
simulations.
It is likely that the approaches in 
Refs.~\cite{Cohen:2003kd,Adams:2004yy,Nagata:2012tc} are closely related.
Now, it is meaningful to consider underlying physics behind the early 
onset of the quark number density by extending the previous approaches.

In this paper, we address the problem of the early onset of 
the quark number density by using the reduction formula. 
According to the previous studies, we introduce 
a physical parametrization of the eigenvalues of the reduced matrix $Q$. 
It will be shown that the parametrization reduces the number density 
operator of the quark to an expression with the Fermi distribution. 
This makes it possible to study the property of the 
quark number density at $T=0$ in a familiar technique. 
The gap of the eigenvalues of $Q$ plays a similar role to a Fermi 
energy in the Fermi distribution. 
The $\mu$-independence of $\det \Delta(\mu)$ at $T=0$ for 
$\mu<m_\pi/2$ is explained by a simple argument based on the 
Fermi distribution with the energy gap. 
We will discuss the origin of the early onset. 

We employ clover-improved Wilson fermions of $N_f=2$. 
The result can be applied to other actions if the reduction formula is 
available.
First we divide the Wilson fermion matrix $\Delta$ into the spatial part and 
temporal part, 
\begin{align}
&\Delta(\mu) =  B -\kappa \left[ e^{+\mu a} (r-\gamma_4) U_4(x) \delta_{x^\prime, x+\hat{4}} \right. \nonumber \\
&\left.+e^{-\mu a} (r+\gamma_4) U^\dagger_4(x^\prime) \delta_{x^\prime, x-\hat{4}}\right], \\
 &B  =  \delta_{x, x^\prime} -\kappa \sum_{i=1}^{3} \Bigl[
(r-\gamma_i) U_i(x) \delta_{x^\prime, x+\hat{i}}  \nn \\
  &+ (r+\gamma_i) U_i^\dagger(x^\prime) \delta_{x^\prime, x-\hat{i}}\Bigr] 
- \kappa  C_{SW} \delta_{x, x^\prime}  \sum_{\mu \le \nu} \sigma_{\mu\nu} 
F_{\mu\nu},
\label{Jul202011eq1}
\end{align}
where $B$ is the spatial Wilson matrix, $\kappa$ the hopping parameter, 
$r$ the Wilson parameter, $C_{SW}$ the clover coefficient. 
We introduce two block-matrices
\begin{subequations}
\begin{align}
\alpha_i =& B^{ab, \mu\sigma}(\vec{x}, \vec{y}, t_i) \; r_{-}^{\sigma\nu} 
         -2  \kappa \; r_{+}^{\mu\nu} \delta^{ab} \delta(\vec{x}-\vec{y}), 
\\
\beta_i =& \Bigl[ B^{ac,\mu\sigma}(\vec{x}, \vec{y}, t_i)\; r_{+}^{\sigma\nu} 
-2 \kappa \; r_{-}^{\mu\nu} \delta(\vec{x}-\vec{y}) 
\Bigr] U_4^{ab}(\vec{y}, t_i).
\label{Eq:2012Feb21eq1}
\end{align}%
\end{subequations}%
$r_\pm = (r \pm \gamma_4)/2$ are projection operators in case of $r=1$, 
which is used in the derivation of the formula.
Using the block matrices, $\det\Delta(\mu)$ is given by~\cite{Nagata:2010xi}
\begin{subequations}
\begin{align}
\det \Delta(\mu) & =\xi^{-\Nred/2}  C_0 \det\left( \xi +  Q \right), 
\label{May1010eq2} \\
Q   &= \prod_{i=1}^{N_t}(\alpha_{i}^{-1} \beta_{i}), \\
C_0 &= \prod_{i = 1}^{N_t} \det(\alpha_i),
\label{Eq:2012Jan01eq4}%
\end{align}%
\label{Eq:2012Jan01eq5}%
\end{subequations}%
where $\xi=e^{-\mu/T}$, $N=4N_c N_s^3 N_t$ and $\Nred = N/N_t$. 
$N$ and $\Nred$ are the dimensions of $\Delta$ and $Q$, respectively. 
$Q$ and $C_0$ are independent of $\mu$.
Calculating $|Q-\lambda I|=0$, we obtain
\begin{align}
\det \Delta (\mu) &=  C_0  \xi^{-\Nred/2}\prod_{n=1}^{\Nred} (\lambda_n + \xi),
\label{Nov292011eq1}
\end{align}
which is the analytic function of $\mu$, and provides values of 
$\det \Delta(\mu)$ for arbitrary $\mu$.

\begin{figure}[htbp] 
\includegraphics[width=6.5cm]{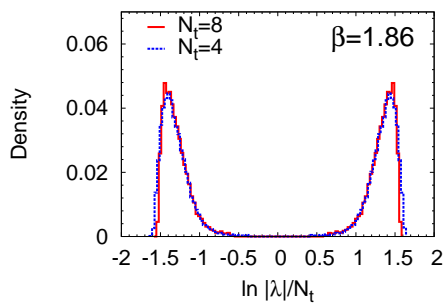}
\caption{
The histogram of the eigenvalue distribution as a function of 
$\ln |\lambda|/N_t$ for $N_t=4$(blue) and $N_t=8$ (red) 
with a fixed scale $a$. The angular part is integrated out.
The agreement between them implies the scaling law $\lambda\sim l^{N_t}$. 
The two peaks are a consequence of the pair nature $\lambda$ and 
$1/\lambda^*$. No eigenvalue exists for $|\lambda|\sim 1$ (gap). 
}
\label{Fig:2012Jan01fig2}
\end{figure}

The reduced matrix $Q$ has three important properties. 
One follows from the $\gamma_5$ hermiticity that the eigenvalues of 
$Q$ appear in pairs $\lambda$ and $1/\lambda^*$, 
which can be seen in a two-peak behavior in Fig.~\ref{Fig:2012Jan01fig2}. 
Accordingly, the eigenvalues are classified into two groups, $|\lambda|<1$ 
and $|\lambda|>1$. 
Here the numerical simulations were done in $N_f=2$ clover-improved 
Wilson fermions and RG-improved gauge action with $8^3\times N_t$ lattice 
and with $m_{\rm ps}/m_{\rm V}=0.8$. Data are taken from Ref.~\cite{Nagata:2012tc}. 
\begin{figure}[htbp]
\includegraphics[width=7cm]{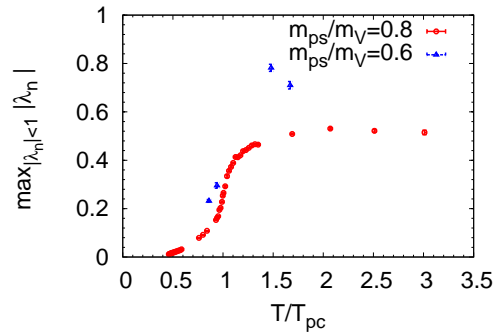}
\caption{
$T$-dependence of the largest eigenvalue in $|\lambda|<1$, i.e., 
$\max_{|\lambda_n|<1} |\lambda_n|$, which is related to 
the pion mass and $\epsilon_{\rm min}$. 
The ensemble average was taken 
over 400 configurations for $m_{\rm PS}/m_{\rm V}=0.8$ 
and 50 configurations for $m_{\rm PS}/m_{\rm V}=0.6$.
}
\label{Fig:2012Jan04fig1}
\end{figure}

Second property is the gap in the eigenspectrum of $Q$.  
No eigenvalue exists in the vicinity of $|\lambda|=1$, see 
Fig.~\ref{Fig:2012Jan04fig1}. 
Gibbs pointed out~\cite{Gibbs:1986hi} that an eigenvalue near the 
gap is related to the pion mass. 
Later Fodor, Szabo and T\^oth investigated further the connection between the 
hadron spectrum and eigenvalues of $Q$~\cite{Fodor:2007ga}. 
In Ref.~\cite{Nagata:2012tc}, we have also calculated the pion mass 
in $N_f=2$ Wilson fermion case. 

Third one is the $N_t$ scaling law. 
Figure~\ref{Fig:2012Jan01fig2} shows the histogram of $\lambda$ 
as a function of $\ln |\lambda|/N_t$ for $N_t=4$ and $N_t=8$ with a 
fixed scale $a$. 
The agreement of the results indicates that an eigenvalue is
parametrized as $|\lambda|\sim l^{N_t}$, where $l$ is a real.  
In Ref.~\cite{Nagata:2012tc}, we derived the low-$T$ limit of 
$\det \Delta(\mu)$ by using these properties of $\lambda$, 
and found that $\det\Delta(\mu)$ is $\mu$-independent for $\mu<m_\pi/2$
in the low $T$ limit.

According to the $N_t$ scaling law, we introduce a parametrization of 
the eigenvalues of $Q$ as $|\lambda|=\exp(-\epsilon/T)$ and 
$|1/\lambda^*|=\exp(\epsilon/T)$, where we describe larger half of the 
eigenvalues in terms of the smaller half of the eigenvalues.  
This parametrization was also implied in Ref.~\cite{Adams:2004yy} and 
suggested from an analogy between $Q$ and the Polyakov 
line~\cite{Nagata:2012tc}. 
The Polyakov loop describes the free energy of static quarks in heavy 
quark limit : $\bra P \ket \sim e^{-F/T}$ with 
$P=\tr \prod_{i=1}^{N_t} U_4(t_i)$. 
Considering an eigenvalue $\forall\lambda, (|\lambda|<1)$, 
the analogy between $P$ and $Q$ may suggest that $\lambda$ is related to 
an energy of a dynamical quark. 
The eigenvalue $\lambda$ has the counterpart $1/\lambda^*$. 
Because the pair nature results from the $\gamma_5$ hermiticity, 
it is natural to identify $1/\lambda^*$ as an energy of an anti-quark.

Adams showed that the reduction formula reproduces 
the fermion determinant obtained from a Matsubara frequency summation method 
in free field case~\cite{Adams:2004yy}. 
We show this correspondence in the present case.
Using the pair nature of the eigenvalues, Eq.~(\ref{Nov292011eq1}) can be rewritten as 
\begin{align}
\det \Delta(\mu) = C_0 \prod_{n=1}^{\Nred/2} (\lambda_n^*)^{-1}(1+\lambda_n\xi^{-1}) (1+\lambda_n^*\xi).
\label{Eq:2012Apr22eq2}
\end{align}
For a while, we drop the phase of $\lambda$. 
If we substitute $\lambda=e^{-\epsilon/T}$ and $\xi=e^{-\mu/T}$, 
then Eq.~(\ref{Eq:2012Apr22eq2}) is reduced to 
\begin{align}
\det \Delta(\mu) = C_0 \prod_{n=1}^{\Nred/2} e^{\epsilon/T} 
(1+e^{-(\epsilon-\mu)/T}) (1+e^{-(\epsilon+\mu)/T}),
\end{align}
which agrees with the Matsubara frequency summation method. 
Note that there is a degeneracy of Dirac components in the free field case. 
This correspondence suggests that $\epsilon a=-\ln |\lambda|/N_t$ 
describes one-particle energy states of a quark for a single configuration. 

Here, we comment on the ensemble average.
The reduction formula provides the fugacity expansion of 
the fermion determinant: $\det \Delta(\mu) 
=\sum c_n \xi^n$, where $c_n$ are functions of 
$\lambda_n$. Taking ensemble average correctly, a canonical partition 
function with a fixed quark number can be obtained, which 
provides a physical free energy of QCD. 
On the other hand, $\epsilon$ corresponds to one-particle energy states 
of a quark for {\it single configuration}. 
This is perhaps related to a ``quasi-energy `` introduced in Ref.~\cite{Cohen:2003kd}, which is eigenvalues of the $\gamma_0$ times Dirac operator.
The ensemble average is crucial to calculate thermodynamical 
quantities, while it is convinient to consider $\epsilon$ 
for the present purpose.

Let us consider the number density operator. 
Equation~(\ref{Eq:2012Apr22eq2}) is rewritten as 
\begin{align}
\det \Delta(\mu) =& C_0 \exp \Biggl( \sum_{n=1}^{\Nred/2} \Bigl[
- \ln \lambda_n^* \nn \\
 &+\ln (1+\lambda_n\xi^{-1} )  +\ln(1+\lambda_n^*\xi)\Bigr] \Biggr).
\label{Eq:2012Apr22eq1}
\end{align}
For large volume, it is reduced to a spectral representation
\begin{align}
\det \Delta (\mu) &= C_0 \exp \Biggl(
4N_c V_s \int_{|\lambda|<1} d\lambda \rho(\lambda) 
\Bigl[ -\ln \lambda^* \nn \\ 
& + \ln (1+\lambda\xi^{-1}) + \ln (
1+\lambda^*\xi)\Bigr]\Biggr).
\label{Eq:2012Mar28eq2}
\end{align}
Here $V_s=N_s^3$, and $\rho(\lambda)$ is the spectral density defined on the 
complex $\lambda$ plane. $\rho$ is normalized by 
\begin{subequations}
\begin{align}
\int d\lambda \rho(\lambda) &=1, \\
\int_{|\lambda|\le 1} d\lambda \rho(\lambda) &= \int_{|\lambda|\ge 1} d\lambda \rho(\lambda) = \frac{1}{2}.
\end{align}
\label{Eq:2012Apr28eq2}
\end{subequations}
Here the normalization conditions are satisfied in any $N_t$.

The number density operator for quarks is given by 
$\hat{n}=[V_s^{-1} T (\del /\del \mu)\det \Delta(\mu) ]/\det \Delta(\mu)$.
We obtain
\begin{align}
\hat{n} = \int_{|\lambda|<1} d \lambda \rho(\lambda) 
\left( 
\frac{\lambda \xi^{-1}}{1+\lambda\xi^{-1}} 
-\frac{\lambda^*\xi}{1+\lambda^*\xi}
\right).
\end{align}
Now, we employ $\lambda=e^{-\epsilon/T+i\theta}$ and $\xi=e^{-\mu/T}$, 
\begin{align}
\hat{n} = \int_{|\lambda|<1} d \lambda \rho(\lambda) 
\left(
\frac{1}{1+e^{(\epsilon-\mu)/T - i\theta}}
-\frac{1}{1+e^{(\epsilon+\mu)/T + i\theta}}
\right),
\end{align}
and
\begin{subequations}
\begin{align}
\int_{|\lambda|<1} d\lambda=
\int_0^\infty \frac{d\epsilon}{T} e^{-2\epsilon/T} \int_{-\pi}^{\pi} d\theta.
\end{align}
\label{Eq:2012Apr28eq1}
\end{subequations}
Then, 
\begin{align}
\hat{n} =& 
\int_0^\infty \frac{d\epsilon}{T} e^{-2\epsilon/T} \int_{-\pi}^{\pi} d\theta
\rho(e^{-\epsilon/T+i\theta}) \nonumber \\
&\times \left(
\frac{1}{1+e^{(\epsilon-\mu)/T - i\theta}}
-\frac{1}{1+e^{(\epsilon+\mu)/T + i\theta}}
\right),
\label{Eq:2012Apr24eq1}
\end{align}
where aside from $\theta$, the first term in the large bracket is the Fermi distribution 
for a quark and the second term is that for an anti-quark.

Once having the Fermi distribution, the zero temperature limit is
obtained from a familiar technique. 
Taking $N_t \to \infty$ with a fixed lattice spacing $a$, 
the Fermi distribution is reduced to the step function. 
It is obvious from $N_t$-independence of Eq.~(\ref{Eq:2012Apr28eq2}) 
that the factor $e^{-2 \epsilon/T}/T$ does not cause any ill-behavior in 
$N_t\to \infty$. 
Then, we obtain
\begin{align}
\hat{n} 
&=\int_0^\mu \frac{d\epsilon}{T} e^{-2\epsilon/T} \int_{-\pi}^{\pi} d\theta
\rho(e^{-\epsilon/T+i \theta}),
\label{Eq:2012Apr23}
\end{align}
where $\mu$ corresponds to a Fermi energy, 
and energy states below $\mu$ are occupied at $T=0$. 

The gap in the eigenspectrum of $Q$ generates the minimum value 
$\epsilon_{\rm min}$ given by 
$\max_{|\lambda|<1}|\lambda|=\exp(-\epsilon_{\rm min}/T)$. 
The average of $\epsilon_{\rm min}$ is shown in Fig.~\ref{Fig:2012Jan04fig1}.
No eigenvalue exists for $\epsilon<\epsilon_{\rm min}$, 
\begin{align}
\rho=0  \;\;\; \mbox{for} \;\;\; \epsilon< \epsilon_{\rm min}.
\end{align}
Hence, we obtain
\begin{align}
\hat{n}(\mu)=0 \;\;\; \mbox{for} \;\;\; \mu<\epsilon_{\rm min}.
\label{Eq:2012Arp24eq1}
\end{align}
Thus, the quark number density is zero at $T=0$ for small $\mu$ 
for any configurations. 
According to Gibbs~\cite{Gibbs:1986hi}, 
$\epsilon_{\rm min}$ corresponds to half the pion mass 
$\epsilon_{\rm min} a= m_\pi a /2$ for each configuration.
An alternative expression was found in Ref.~\cite{Fodor:2007ga}, 
where the expression was given for the ensemble average. 
Although there is a difference between the results in Ref.~\cite{Gibbs:1986hi} 
and \cite{Fodor:2007ga}, it is expected~\cite{Fodor:2007ga} that the results
become the same for large temporal lattice size. 
Hence, we employ $\epsilon_{\rm min}=m_\pi/2$. 
$\hat{n}$ is zero for configuration by configuration for 
$\mu<\epsilon_{\rm min}$, and therefore its average is also zero, 
$\bra \hat{n}\ket=0$.

Now, we discuss the onset of the quark number density.
A phenomenological expectation is that the quark number 
density starts to differ from zero at $\mu=M_N/3$. 
This expectation is based on the fact that quarks are confined 
inside hadrons. 
The present result, e.g. Eq.~(\ref{Eq:2012Apr24eq1}) indicates 
that the introduction of the quark chemical potential actually affects
an excitation of quarks even in the confinement phase at 
least for single configuration. 
For $\mu<\epsilon_{\rm min}$, $\mu$ is not enough to excite quarks 
for any configurations, and the quark number density is zero. 
Once $\mu$ exceeds $\epsilon_{\rm min}$, the chemical potential causes 
excitation of quarks and leads to nonzero quark number density 
for each configuration. This is the origin of the early 
onset of the quark number density.

However, it does not necessarily mean that the average of 
the quark number density is nonzero.
Once $\hat{n}$ takes nonzero values, the phase of $\hat{n}$ also 
becomes nonzero. It is known that the sign problem becomes severe if 
$\mu$ goes beyond $m_\pi/2$~\cite{Nagata:2012tc}. 
Hence, it is unclear what happens for $m_\pi/2<\mu<M_N/3$. 
The complex phase of $\hat{n}$ causes the cancellation of 
the quark number density over configurations, 
which may lead to $\bra \hat{n}\ket=0$ for $\mu<M_N/3$. 

Another possibility is that finite density effects change 
$\epsilon_{\rm min}$. 
The relation $\epsilon_{\rm min}=m_\pi/2$ is based on the fact that 
the pion is the lightest hadron.
Although QCD inequalities do not hold for finite density region, 
the pion is the lightest hadron at least for nuclear matter density. 
Hence the relation would hold at least for nuclear matter density. 
It seems that this possibility is unlikely.
If $\bra \hat{n}\ket=0$ for $m_\pi/2<\mu<M_N/3$, then it would be
a consequence of the cancellation of the quark number 
density over configurations.

In summary, we have addressed the longstanding problem of the 
onset of the quark number density. 
Based on previous studies, we considered it by using the properties of the 
reduction formula obtained from the lattice QCD simulations.
We parametrized the eigenvalues of the reduction formula based on 
some properties of the eigenvalues. 
The parametrization reduces the quark number density operator 
to the form with the Fermi distribution of a quark, where 
the eigenvalues of the reduced matrix correspond to 
one-particle energy states of a quark at single configuration level. 
Then, the $\mu$-independence of the fermion determinant is 
explained by the familiar technique. 
The present result shows that the quark chemical potential actually 
affects an excitation of quarks.
Small $\mu$ can not excite any quark at $T=0$, which causes 
the $\mu$-independence of the fermion determinant at $T=0$. 
On the other hand, if $\mu$ exceeds a certain value related to 
the pion mass, it can excite quarks even in the confinement phase at 
least one-configuration level, which is the origin of the early onset of 
the quark number density. 
If the quark number density remains zero up to $\mu=M_N/3$, 
the complex phase of the quark number density would cause the cancellation
over configurations. 
This result provides a hint to go beyond $\mu=m_\pi/2$ in lattice simulations.

For further confirmation, the spectral properties of the reduced matrix 
should be investigated in future lattice simulations for smaller quark mass, 
and larger lattice. Particularly, the behavior of the gap and $N_t$-dependence 
are important. 
It is also important to study the onset of the baryon number density 
 by both theoretical and experimental studies. 
Our result suggests the validity of the phase quench QCD and the orbifold 
equivalence  at low temperature for 
$\mu<m_\pi/2$~\cite{Cherman:2010jj,Hanada:2011ju}. 
Phase quench simulations are also interesting for small $\mu$.

We especially thank to A. Nakamura for valuable comments and interesting
discussions. 
We also thank to Ph. de Forcrand, M. Hanada, S. Hashimoto, S. Motoki, Y. Nakagawa and T. Saito for interesting discussions.
KN also thank to T. Misumi, A. Ohnishi for hospitality during the 
NTFL workshop.
This work was supported by Grants-in-Aid for Scientific Research 20340055, 20105003, 23654092. 
The simulation was performed on NEC SX-8R at RCNP, NEC SX-9 at CMC, Osaka University.


\begin{thebibliography}{50}

\bibitem{Aoki:2006we}
Y.~Aoki, G.~Endrodi, Z.~Fodor, S.~Katz, and K.~Szabo,
\newblock Nature {\bf 443}, 675 (2006), arXiv:hep-lat/0611014.

\bibitem{Borsanyi:2010bp}
Wuppertal-Budapest Collaboration, S.~Borsanyi {\em et~al.},
\newblock JHEP {\bf 1009}, 073 (2010), arXiv:1005.3508.

\bibitem{Borsanyi:2010cj}
S.~Borsanyi {\em et~al.},
\newblock JHEP {\bf 1011}, 077 (2010), arXiv:1007.2580.

\bibitem{Bazavov:2012ty}
for HotQCD collaboration, A.~Bazavov,
\newblock (2012), arXiv:1201.5345.

\bibitem{Stephanov:2007fk}
M.~Stephanov,
\newblock PoS {\bf LAT2006}, 024 (2006), arXiv:hep-lat/0701002.

\bibitem{Philipsen:2008gf}
O.~Philipsen,
\newblock Prog.Theor.Phys.Suppl. {\bf 174}, 206 (2008), arXiv:0808.0672.

\bibitem{Fukushima:2010bq}
K.~Fukushima and T.~Hatsuda,
\newblock Rept.Prog.Phys. {\bf 74}, 014001 (2011), arXiv:1005.4814.

\bibitem{Ohnishi:2011aa}
A.~Ohnishi,
\newblock (2011), arXiv:1112.3210.

\bibitem{Demorest:2010bx}
P.~Demorest, T.~Pennucci, S.~Ransom, M.~Roberts, and J.~Hessels,
\newblock Nature {\bf 467}, 1081 (2010), arXiv:1010.5788.

\bibitem{Barbour:1999mc}
I.~Barbour, S.~Hands, J.~B. Kogut, M.-P. Lombardo, and S.~Morrison,
\newblock Nucl. Phys. {\bf B557}, 327 (1999), arXiv:hep-lat/9902033.

\bibitem{Barbour:1997ej}
I.~M. Barbour, S.~E. Morrison, E.~G. Klepfish, J.~B. Kogut, and M.-P. Lombardo,
\newblock Nucl. Phys. Proc. Suppl. {\bf 60A}, 220 (1998),
  arXiv:hep-lat/9705042.

\bibitem{Barbour:1997bh}
I.~M. Barbour, S.~E. Morrison, E.~G. Klepfish, J.~B. Kogut, and M.-P. Lombardo,
\newblock Phys. Rev. {\bf D56}, 7063 (1997), arXiv:hep-lat/9705038.

\bibitem{Barbour:1991vs}
I.~M. Barbour and A.~J. Bell,
\newblock Nucl. Phys. {\bf B372}, 385 (1992).

\bibitem{Aloisio:1998cn}
R.~Aloisio, V.~Azcoiti, G.~Di~Carlo, A.~Galante, and A.~F. Grillo,
\newblock Phys. Lett. {\bf B428}, 166 (1998), arXiv:hep-lat/9802004.

\bibitem{Crompton:2001ws}
P.~R. Crompton,
\newblock Nucl. Phys. {\bf B619}, 499 (2001), arXiv:hep-lat/0108016.

\bibitem{Cohen:2003kd}
T.~D.~. Cohen,
\newblock Phys.Rev.Lett. {\bf 91}, 222001 (2003), arXiv:hep-ph/0307089.

\bibitem{Adams:2004yy}
D.~H. Adams,
\newblock Phys.Rev. {\bf D70}, 045002 (2004), arXiv:hep-th/0401132.

\bibitem{Nagata:2012tc}
XQCD-J Collaboration, K.~Nagata, S.~Motoki, Y.~Nakagawa, A.~Nakamura, and
  T.~Saito,
\newblock (2012), arXiv:1204.1412.

\bibitem{Nagata:2010xi}
K.~Nagata and A.~Nakamura,
\newblock Phys.Rev. {\bf D82}, 094027 (2010), arXiv:1009.2149.

\bibitem{Gibbs:1986hi}
P.~E. Gibbs,
\newblock Phys. Lett. {\bf B172}, 53 (1986).

\bibitem{Fodor:2007ga}
Z.~Fodor, K.~Szabo, and B.~Toth,
\newblock JHEP {\bf 0708}, 092 (2007), arXiv:0704.2382.

\bibitem{Cherman:2010jj}
A.~Cherman, M.~Hanada, and D.~Robles-Llana,
\newblock Phys.Rev.Lett. {\bf 106}, 091603 (2011), arXiv:1009.1623.

\bibitem{Hanada:2011ju}
M.~Hanada and N.~Yamamoto,
\newblock JHEP {\bf 1202}, 138 (2012), arXiv:1103.5480.

\end{thebibliography}

\end{document}